 \documentclass[final,1p,times]{elsarticle}

 \usepackage{graphicx}
\usepackage{amssymb}

\newcommand{\bea}{\begin{eqnarray}}
\newcommand{\eea}{\end{eqnarray}}
\newcommand{\beaa}{\begin{eqnarray*}}
\newcommand{\eeaa}{\end{eqnarray*}}
\newcommand{\be}{\begin{equation}}
\newcommand{\ee}{\end{equation}}
\newcommand{\bear}{\begin{array}}
\newcommand{\eear}{\end{array}}

\newcommand{\eps}{\varepsilon}

\newcommand{\dtwo}[2]{\frac{\partial^2 #1}{\partial #2^2}}
\newcommand{\done}[2]{\frac{\partial #1}{\partial #2}}

\begin{document}

\begin{frontmatter}



\title{Piezoelectric resonance in Rochelle
salt: the contribution of diagonal strains}


\author{A.P.Moina}

\address{Institute for Condensed Matter Physics, 1 Svientsitskii Street,
79011, Lviv, Ukraine}

\begin{abstract}
Within the framework of two-sublattice Mitsui model with taking
into account the shear strain $\varepsilon_4$ and the diagonal
strains $\eps_2$ and $\eps_3$, a dynamic dielectric response of
Rochelle salt X-cuts is considered. Experimentally observed
phenomena of crystal clamping by high frequency electric field,
piezoelectric resonance and microwave dispersion are described. It
is shown that the lowest resonant frequency is always associated
with the $\eps_4$ shear mode

\end{abstract}


\end{frontmatter}


\section{Introduction}
\label{sect1} Crystals of Rochelle salt have been attracting an
interest of physicists due to their practical applications in
past, and now from the fundamental point of view mostly. In
contrast to most of the known ferroelectrics, in Rochelle salt the
ferroelectric phase exists only in a temperature interval between
two second order phase transitions at 255 and 297~K. Spontaneous
polarization $P_1$ is accompanied by shear strain $\eps_4$. The
ferroelectric phase is monoclinic (P$2_111$); both paraelectric
phases are orthorhombic (P$2_12_12_1$); all phases are
non-centrosymmetric and piezoelectric.

Dynamical dielectric response of Rochelle salt exhibits several
dispersion regions. Those are related to the domain wall motion
\cite{Araujo,shylnikov,Poplavko1} (below 1~kHz), piezoelectric
resonance \cite{Poplavko1,Leonovici,muller} (between 10~kHz and
10~MHz), microwave relaxation \cite{int4}, and the submillimeter
(100-700~GHz) resonances \cite{Volkov}. While the very low and
very high frequency dispersions are relatively well studied, the
presence of the piezoelectric resonance dispersion in Rochelle
salt is acknowledged at best. Quantitative data is available
mostly for the temperature dependence of the first resonance
frequency \cite{Leonovici,Mason,muller}, whereas very little
information has been  obtained  \cite{Leonovici} about details of
the temperature or frequency dependence of the dielectric
permittivity, mode assignment for different crystal cuts, etc.

Behavior of Rochelle salt is usually described within a
two-sublattice Ising model with an asymmetric double-well
potential (Mitsui model \cite{int3,83,86}). The pseudospin
dynamics is considered within the Bloch equations or Glauber
approach methods.

Rochelle salt is a perfect example of a system, whose dynamic
dielectric response cannot be correctly described without taking
into account the deformational effects. Their influence is
revealed, in particular, in the phenomena of piezoelectric
resonance and crystal clamping by a high-frequency measuring
field, none of which can be obtained within  theories based on
underformable versions of the Mitsui model
\cite{83,our-comp-piezo}. Such theories yield a diverging
relaxation time at the Curie point and, as a result, incorrect
temperature behavior of the microwave permittivity near the phase
transitions.

In \cite{ourrs2}  the dynamic dielectric permittivity of Rochelle
salt has been calculated, using the model with piezoelectric
coupling \cite{ourrs} with the shear strain $\eps_4$, for the
entire frequency range from the static limit (in the ferroelectric
phase from about 1~kHz) to 10$^{11}$~Hz, including  the
piezoelectric resonance region. For a coupled dynamics of the
shear strain $\eps_4$ -- pseudospin system, the standard methods
of description of the lattice strain dynamics \cite{Mason} based
on Newtonian equations of motion has been combined with the
Glauber approach to pseudospin dynamics. Evolution of the
dielectric permittivity from the static free crystal value via the
piezoelectric resonances to the clamped crystal value and to the
microwave relaxation has been described. Recently, in
\cite{Andrusyk} it has been pointed out  that boundary conditions
in \cite{ourrs2} were not set correctly, which resulted in the
underestimated values of the resonant frequencies; a correct
equation for the resonant frequencies related to the shear
$\eps_4$ mode has been obtained \cite{Andrusyk}.

In the paraelectric phases in Rochelle salt the longitudinal field
$E_1$ excites only the shear mode $\eps_4$, as $d_{14}$ is there
the only non-zero piezoelectric coefficient associated with $E_1$.
However, in the ferroelectric phase it can also excite the
extensional modes associated with the diagonal strains $\eps_1$,
$\eps_2$, and $\eps_3$ via the non-zero coefficients $d_{11}$,
$d_{12}$, $d_{13}$. The contributions of the extensional modes to
the dielectric permittivity of Rochelle salt at frequencies far
from the piezoelectric resonance region are not expected to be
crucial, due to smallness of $d_{1i}$ ($i=1,2,3$) in comparison
with $d_{14}$. The presence of these modes, however, changes the
permittivity in the resonance region, at least giving rise to
additional resonance peaks, and this should be explored.

In this paper we follow the same approach that has been used
previously \cite{ourrs2} for the case with only one shear mode.
The modification of the Mitsui model of \cite{our-diagonal} that
takes into account both the shear strain $\eps_4$ and the diagonal
strains is exploited. The drawbacks of the previous calculations
related to setting the boundary conditions are removed. The
expression for the dynamic dielectric permittivity and equations
for the resonant frequencies of Rochelle salt X-cuts is obtained.

\section{System thermodynamics in presence of diagonal strains}

For the sake of the  reader's convenience, we shall present here
the expressions for the related to the shear strain $\eps_4$
thermodynamic and physical characteristics of Rochelle salt
obtained  within the modified two-sublattice Mitsui model with the
shear strain $\eps_4$ and with the diagonal strains
\cite{our-diagonal}.

The system behavior is described in terms of the following linear
combinations of the mean  values of the pseudospins belonging to
different sublattices
\[
\xi = \frac12(\langle\sigma_{q1}\rangle +
\langle\sigma_{q2}\rangle), \quad \sigma =
\frac12(\langle\sigma_{q1}\rangle - \langle\sigma_{q2}\rangle);
\]
$\xi$ is the parameter of ferroelectric ordering in the system.

The thermodynamic potential of the  model \cite{our-diagonal}
within the mean field approximation reads
 \begin{equation}
 \label{pot}
 g_{2E}(T) =U_{ seed} + \frac{J +K}4\xi^2 + \frac{J
-K}4\sigma^2
 - \frac{2\ln 2}{\beta} -  \frac1\beta\ln \cosh
\frac{\gamma + \delta}2
 \cosh \frac{\gamma - \delta}2,
\end{equation}
where ${U_{seed}}$ is the phenomenological part of the
thermodynamical potential, representing the energy of the host
lattice of heavy ions which forms the asymmetric double-well
potentials for the pseudospins (see \cite{our-diagonal});
$\beta=1/k_{B}T$, $k_B$ is the Boltzmann constant, and
\[
\gamma = \beta\left( \frac{J + K}{2}\xi - 2\psi_4\varepsilon_4 +
\mu_1E_1\right), \quad \delta = \beta \left( \frac{J - K}{2}\sigma
+ \Delta\right).
\]
Here $\mu_1$ is the effective dipole moment. The model parameter
$\psi_4$ describes the internal field created by the piezoelectric
coupling with $\varepsilon_4$. It is assumed that a longitudinal
electric field $E_1$ is applied.

The parameters $J$, $K$ are the Fourier-transforms (at ${\bf
k}=0$) of the constants of interaction between  pseudospins
belonging to the same and to different sublattices, respectively.
They, along with  the double well potential asymmetry parameter
$\Delta$, are assumed \cite{our-diagonal} to be linear functions
of the diagonal strains
\begin{equation}\label{2.2}
    J \pm K= J_0 \pm K_0+2 \sum\limits_{i = 1}^3 {\psi_{i}^\pm \varepsilon _i
    },\quad \Delta = \Delta _0 + \sum\limits_{i = 1}^3 {\psi_{3i}
\varepsilon _i }.
\end{equation}

The stress-strain relations and polarization have been obtained in
the following form \cite{our-diagonal}
\begin{eqnarray}
\label{2.4} && \sigma_i = \sum\limits_{j = 1}^3 c_{ij}^{E0}
[\varepsilon _j -\alpha_j^0(T-T_j^0)]
- \frac{1}{2v}\psi_i^+  \xi ^2 - \frac{1}{2v}\psi_i^- \sigma ^2
- \frac{1}{v}\psi_{3i} \sigma, \quad ({i } = 1-3)\nonumber \\
\label{2.4a} && \sigma_4 = c_{44}^{E0} \varepsilon _4 - e_{14}^0
E_1 +
2\frac{\psi _4 }{v}\xi , \\
 \label{2.4b} && P_1 = e_{14}^0 \varepsilon
_4 + \chi _{11}^{\varepsilon 0} E_1 + \frac{\mu _1 }{v}\xi .
 \end{eqnarray}
 $\sigma_i$ are the components of the stress tensor in Voigt
notations; $c_{ij}^{E0}$ are the ``seed'' elastic constants;
$\alpha_i^0$ are the ``seed'' thermal expansion coefficients.

\section{Vibrations of X-cuts of Rochelle salt}
We  consider vibrations of a thin rectangular  $L_y\times L_z$
plate of a Rochelle salt crystal cut in the (100) plane (X-cut)
induced by time-dependent electric field $E_1=E_{1t}\exp(i\omega
t)$. This field gives rise to the shear strain $\eps_4$ at all
temperatures, as well as to the diagonal strains $\eps_1$,
$\eps_2$, $\eps_3$ in the ferroelectric phase. We take into
account the in-plane vibrational modes, allowed by the system
symmetry, and neglect the out-of-plane mode associated with
$\eps_1$.

Dynamics of the pseudospin subsystem will be described within the
Glauber approach \cite{int9}, where the  kinetic equations for the
time-dependent averages $\xi$  and $\sigma$ have the form
\cite{ourrs}
\begin{eqnarray}
-\alpha\frac{d}{dt}\xi=\xi-\frac{1}{2}[\tanh\frac12(\gamma+\delta)+
\tanh\frac12(\gamma-\delta)],
\nonumber\\
\label{3.3}
-\alpha\frac{d}{dt}\sigma=\sigma-\frac{1}{2}[\tanh\frac12(\gamma+\delta)-
\tanh\frac12(\gamma-\delta)].
\end{eqnarray}
Here $\alpha$  is the parameter setting the scale of the dynamic
processes in the pseudospin subsystem.  The form of (\ref{3.3}) is
not affected by inclusion of the diagonal strains into
consideration.

Dynamics of the strains  will be described by the standard method,
using  classical  (Newtonian) equations of motion \cite{Masonbook}
of an elementary volume
\begin{equation}
\label{11} \rho\dtwo{\eta_i}{t}=\sum_k \done{\sigma_{ik}}{x_k},
\end{equation}
where $\rho=1.767$~g/cm$^3$ is the crystal density, $\eta_i$ are
the displacements of an elementary volume along the axis $x_i$,
$\sigma_{ik}$ are components of the the mechanical stress tensor.
Relevant to our case are the displacements $\eta_2$ and $\eta_3$,
giving the strains
\[
\eps_2= \done{\eta_2}{y},\quad \eps_3= \done{\eta_3}{z},\quad
\eps_4=\done{\eta_2}{z}+\done{\eta_3}{y}.
\]

At small deviations from the equilibrium the dynamic variables
$\xi$, $\sigma$, and $\eps_j$ (or $\eta_i$) can be presented as
sums of the equilibrium values  and of the fluctuational
deviations, while the deviations are taken to be in the form of
harmonic waves
 \[
\xi=\tilde\xi+\xi_t(y,z)\exp(i\omega t), \quad
\sigma=\tilde\sigma+\sigma_t(y,z)\exp(i\omega t), \quad
\eta_i=\tilde\eta_i+\eta_{it}(y,z)\exp(i\omega t).
\]
Equations (\ref{2.4})-(\ref{2.4b}), (\ref{3.3}), and (\ref{11})
can be expanded in terms of these deviations up to the linear
terms. For $\tilde\xi$ and $\tilde\sigma$ we obtain the same
equations that follow from the condition of the thermodynamic
potential (\ref{pot}) extremum \cite{our-diagonal}.

From (\ref{2.4})-(\ref{2.4b}) we get the following constitutive
equations
\begin{eqnarray}
\label{2.4f} && \sigma_i(y,z) = \sum\limits_{j = 1}^3 c_{ij}^{E0}
\varepsilon _{jt}(y,z)
 - \frac{1}{v}\psi_i^+  \tilde\xi\xi_t(y,z) - \frac{1}{v}\psi_i^-
\tilde\sigma\sigma_t(y,z)
- \frac{1}{v}\psi_{3i} \sigma_t(y,z), \nonumber\\
\label{2.4af} && \sigma_4(y,z) = c_{44}^{E0} \varepsilon
_{4t}(y,z) - e_{14}^0 E_{1t} +
2\frac{\psi _4 }{v}\xi_t(y,z) , \\
 \label{2.4bf} && P_{1t}(y,z) = e_{14}^0 \varepsilon
_{4t}(y,z) + \chi _{11}^{\varepsilon 0} E_{1t} + \frac{\mu _1
}{v}\xi_t(y,z).
 \end{eqnarray}
(for $i  = 1-3$). Taking these into into account, from
(\ref{3.3}), and (\ref{11}) we get the following system of
equations
 \bea \label{eta}
&&-\alpha\frac{d}{dt}\xi_t+a_{1}\xi_t+a_{2}\sigma_t
+a_{32}\eps_{2t}+a_{33}\eps_{3t}+a_{34}\eps_{4t} =a_{02}E_{1t},
\nonumber\\
 &&-\alpha\frac{d}{dt}\sigma_t+b_{1}\xi_t+b_{2}\sigma_t
+b_{32}\eps_{2t}+b_{33}\eps_{3t}+b_{34}\eps_{4t} =b_{02}E_{1t},\nonumber\\
&&\rho\dtwo{\eta_{2t}}{t}=c_{22}^{E0}\dtwo{\eta_{2t}}{y}+
c_{23}^{E0}\frac{\partial^2\eta_{2t}}{\partial y\partial z}+
c_{44}^{E0}(\dtwo{\eta_{2t}}{z}+\frac{\partial^2\eta_{3t}}{\partial
y\partial z})+\nonumber\\
&&\quad {} + \frac{2\psi_4}{ v}\done{\xi_t}{z}-
\frac1v[\psi^+_2\tilde\xi \done{\xi_t}{y}+(\psi_2^-\tilde\sigma+\psi_{32})\done{\sigma_t}{y}], \nonumber\\
\label{system}
&&\rho\dtwo{\eta_{3t}}{t}=c_{23}^{E0}\frac{\partial^2\eta_{2t}}{\partial
y\partial z}+ c_{33}^{E0}\frac{\partial^2\eta_{3t}}{\partial z^2}+
c_{44}^{E0}(\dtwo{\eta_{3t}}{y}+\frac{\partial^2\eta_{2t}}{\partial
y\partial z})+\\
&&\quad {} + \frac{2\psi_4}{ v}\done{\xi_t}{y}-
\frac1v[\psi^+_3\tilde\xi
\done{\xi_t}{y}+(\psi_3^-\tilde\sigma+\psi_{33})\done{\sigma_t}{z}],\nonumber
 \eea
with \beaa &&a_{1}=-1+\beta\frac{ J+K}{4} \lambda_1, \quad
  a_{2}=\beta\frac{ K-  J}{4}\lambda_2,\\
&&
a_{32,3}=\frac{\beta}{2}[\tilde\xi\lambda_1\psi^+_{2,3}-\lambda_2(\psi^-_{2,3}\tilde\sigma+\psi_{32,3})],\quad
a_{34}=-\beta\psi_4\lambda_1,\quad
 a_{02}=-\frac{\beta\mu_1}{2}\lambda_1, \\
&&b_{1}=-\beta\frac{ J+ K}{4}\lambda_2,\quad
  b_{2}=-1+\beta\frac{J-K}{4}\lambda_2, \quad
\\
&&
b_{32,3}=-\frac{\beta}{2}[\tilde\xi\lambda_2\psi^+_{2,3}-\lambda_1(\psi^-_{2,3}\tilde\sigma+\psi_{32,3})],\quad
b_{34}=\beta\psi_4\lambda_2,\quad
  b_{02}=\frac{\beta\mu_1}{2}\lambda_2.
  \eeaa
(Hereafter it is implied that the  deviations are functions of $y$
and $z$.) We introduced the following notations
\[
\lambda _1 = 1 - \xi ^2 - \sigma ^2, \quad \lambda _2 = 2\xi
\sigma.
\]

Solving the two first equations of (\ref{system}) with respect to
$\xi_t$ at $\eta_{2t}=\eta_{3t}=0$ (regime of a mechanically
clamped crystal), substituting the result into (\ref{2.4bf}), and
differentiating that with respect to the field, we find the
dynamic dielectric permittivity of a clamped crystal
\begin{equation} \label{chi_clamped}
\eps_{11}^\eps(\omega)=\eps_{11}^{\eps0}+4\pi\frac{\beta\mu_1^2}{2v}F_1(\alpha\omega),\quad
\eps_{11}^{\varepsilon 0}=1+4\pi\chi_{11}^{\varepsilon 0}.
\end{equation}
where
\begin{eqnarray*}
&&F_1(\alpha\omega)=\frac{i\alpha\omega \lambda_1 +
\varphi_3}{(i\alpha\omega)^2 + (i\alpha\omega) \varphi_1
+\varphi_2 },\quad \varphi_1=2-\frac{\beta J}{2}\lambda_1,\\
&& \varphi _2 = 1 - \frac{\beta J}{2}\lambda _1 - \beta
^2\frac{K^2 - J^2}{16}(\lambda _1^2 - \lambda _2^2 ),\quad \varphi
_3 = \lambda _1 + \frac{\beta (K - J)}{4}(\lambda _1^2 - \lambda
_2^2 ).
\end{eqnarray*}
 This is the same expression that has been obtained previously
\cite{ourrs2} for the Mitsui model with the shear strain $\eps_4$,
but without the diagonal strains.

In the regime of a mechanically free crystal, the two first
equations of (\ref{system}) give
\begin{eqnarray}
&&\xi_t(y,z)=
\frac{\beta\mu_1}{2}F_1(\alpha\omega)E_{1t}-\beta\psi_4
F_1(\alpha\omega)\eps_{4t}(y,z)-\nonumber\\
&&\quad{}-\frac{\beta}{2}F_{42}^\xi(\alpha\omega)\eps_{2t}(y,z)
-\frac{\beta}{2}F_{43}^\xi(\alpha\omega)\eps_{3t}(y,z)
,\nonumber\\
&& \label{xi-sigma-eps} \sigma_t(y,z)=
\frac{\beta\mu_1}{2}F_1^\sigma(\alpha\omega)E_{1t}-\beta\psi_4
F_1^\sigma(\alpha\omega)\eps_{4t}(y,z)-\nonumber\\
&&\quad{}-\frac{\beta}{2}F_{42}^\sigma(\alpha\omega)\eps_{2t}(y,z)
-\frac{\beta}{2}F_{43}^\sigma(\alpha\omega)\eps_{3t}(y,z),
\end{eqnarray}
where
\begin{eqnarray*}
&&F_{4j}^\xi(\alpha\omega)=\frac{\varphi_{4j}+
(i\alpha\omega)[\tilde\xi\lambda_1\psi_j^+-(\psi_j^-\tilde\sigma+\psi_{3j})\lambda_2]}{D(\omega)},\nonumber\\
&&F_{4j}^\sigma(\alpha\omega)=-\frac{-\tilde\xi\lambda_1\psi_j^++(\psi_j^-\tilde\sigma+\psi_{3j})\varphi_5+
(i\alpha\omega)[\tilde\xi\lambda_2\psi_j^+-(\psi_j^-\tilde\sigma+\psi_{3j})\lambda_1]}{D(\omega)},\nonumber\\
&&F_{1}^\sigma(\alpha\omega)=-\frac{\lambda_2(1+i\alpha\omega)}{D(\omega)},\nonumber\\
&&D(\omega)=(i\alpha\omega)^2 + (i\alpha\omega) \varphi_1
+\varphi_2,\\
&&\varphi _{4i} = \psi_i^+ \xi \varphi _3 - \left( {\psi_i^-
\sigma + \psi_{3i} } \right)\lambda _2 ,\quad \varphi _5 = \lambda
_1 - \frac{\beta (K + J)}{4}(\lambda _1^2 - \lambda _2^2 ).
\end{eqnarray*}
 Hence, the field-dependent part of polarization is
\begin{equation} P_{1t}(y,z)=
\left[\chi_{11}^{\eps0}+\frac{\beta\mu_1^2}{2v}F_1(\alpha\omega)\right]E_{1t}+\sum_{i=2,3,4}e_{1i}(\omega)\eps_{it}(y,z),
\end{equation}
where
\begin{equation}
\label{e_om}
e_{12,3}(\omega)=\frac{\beta\mu_1}{2v}F_{42,3}^\xi(\alpha\omega),\quad
e_{14}(\omega)=e_{14}^{0}-
\frac{\beta\mu_1\psi_4}{v}F_1(\alpha\omega) \end{equation} are
dynamic piezoelectric coefficients.

 The observable
dynamic dielectric permittivity at constant stress
$\chi_{11}^\sigma(\omega)$ is expressed via the derivative from
the polarization averaged over the sample volume
\begin{eqnarray}
&&\eps_{11}^\sigma(\omega)=1+\frac{4\pi}{L_yL_z}\frac{\partial{}}{\partial
E_{1t}} \int_0^{L_y} dy\int_0^{L_z} dz P_{1t}(y,z)=\nonumber\\
&&\quad=\eps_{11}^\eps(\omega)+
4\pi\sum_{i=2,3,4}e_{1i}(\omega)\done{}{E_{1t}}\overline{\eps_{it}(y,z)},
\end{eqnarray}
where
\[
\overline{\eps_{it}}=\frac{1}{L_yL_z}\int_0^{L_y}dy\int_0^{L_z}dz
\eps_{it}(y,z).
\]
The remaining problem is to find $\eps_{it}(y,z)$.

Taking into account (\ref{xi-sigma-eps}), from the two last
equation of (\ref{system}) it follows that
 \begin{eqnarray}
&&-\rho\omega^2\eta_{2t}=\tilde{c}_{22}^{E}(\omega)\dtwo{\eta_{2t}}{y}
+2\tilde{c}_{24}^{E}(\omega)\frac{\partial^2\eta_{2t}}{\partial
y\partial z}+
\tilde{c}_{44}^{E}(\omega)\dtwo{\eta_{2t}}{z}+\nonumber\\
&&\qquad{}+\tilde{c}_{24}^{E}(\omega)\dtwo{\eta_{3t}}{z}+
[\tilde{c}_{44}^{E}(\omega)+\tilde{c}_{23}^{E}(\omega)]\frac{\partial^2\eta_{3t}}{\partial
y\partial z}+ \tilde{c}_{34}^{E}(\omega)\dtwo{\eta_{3t}}{z}
 ,\nonumber\\
&&-\rho\omega^2\eta_{3t}=\tilde{c}_{24}^{E}(\omega)\dtwo{\eta_{2t}}{y}
+[\tilde{c}_{23}^{E}(\omega)+\tilde{c}_{44}^{E}(\omega)]\frac{\partial^2\eta_{2t}}{\partial
y\partial z}+
\tilde{c}_{34}^{E}(\omega)\dtwo{\eta_{2t}}{z}+\nonumber\\
&&\qquad{}+\tilde{c}_{44}^{E}(\omega)\dtwo{\eta_{3t}}{z}+
2\tilde{c}_{34}^{E}(\omega)\frac{\partial^2\eta_{3t}}{\partial
y\partial z}+ \tilde{c}_{33}^{E}(\omega)\dtwo{\eta_{3t}}{z},
\label{Christoffel}
\end{eqnarray}
which is nothing but the Christoffel equations, with the
frequency-dependent elastic constants $\tilde c_{ij}^E(\omega)$
given by the model expressions
 \begin{eqnarray}
 && \tilde
c_{i4}^E(\omega)=\frac{\beta\psi_4}{v}F_{4i}^\xi(\alpha\omega),\nonumber\\
&& \tilde c_{ij}^E(\omega)=c_{ij}^{E0}-\frac{\beta} {2v}\Bigl\{
\psi_j^+\tilde\xi F_{2i}^\xi(\alpha\omega)+ ( \psi_j^- \sigma +
\psi_{3j} )F_{2i}^\sigma(\alpha\omega) \Bigr\},
 \nonumber\\
&&\label{c_om}
\tilde{c}_{44}^{E}(\omega)=c_{44}^{E0}-\frac{2\beta\psi_4^2}{v}F_1(\alpha\omega).
\end{eqnarray}
Their  frequency variation is perceptible only in the  region of
the microwave dispersion of the dielectric susceptibility.
However, in the piezoelectric resonance region, which is expected
to be in the $10^4-10^7$~Hz range, depending on temperature and
sample dimensions, well separated from the region of the microwave
relaxation, $\tilde{c}_{ij}^E(\omega)$ and $e_{ij}(\omega)$
(\ref{e_om}) are very close to the corresponding static quantities
(see \cite{our-diagonal,our-comp}) and coincide with them at
$\omega\to0$.

We can rewrite (\ref{Christoffel}) in a different form, where it
will be more convenient to set the boundary conditions. We
differentiate (\ref{Christoffel}) with respect to $y$ and $z$,
transforming it into three equations for the strains $\eps_{2t}$,
$\eps_{3t}$, $\eps_{4t}$ instead of the displacements $\eta_{2t}$
and $\eta_{3t}$
\begin{eqnarray}
&&-\rho\omega^2\eps_{2t}=\tilde{c}_{22}^{E}(\omega)\dtwo{\eps_{2t}}{y}
+\tilde{c}_{24}^{E}(\omega)\frac{\partial^2\eps_{2t}}{\partial
y\partial z}+
\tilde{c}_{44}^{E}(\omega)\dtwo{\eps_{2t}}{z}+\nonumber\\
&&\qquad{}+\tilde{c}_{24}^{E}(\omega)\dtwo{\eps_{4t}}{y}+
[\tilde{c}_{44}^{E}(\omega)+\tilde{c}_{23}^{E}(\omega)]\frac{\partial^2\eps_{3t}}{\partial
y^2}+
\tilde{c}_{34}^{E}(\omega)\frac{\partial^2\eps_{3t}}{\partial
y\partial z}
 ,\nonumber\\
&&-\rho\omega^2\eps_{3t}=\tilde{c}_{24}^{E}(\omega)\frac{\partial^2\eps_{2t}}{\partial
y\partial z}
+[\tilde{c}_{23}^{E}(\omega)+\tilde{c}_{44}^{E}(\omega)]\dtwo{\eps_{2t}}{z}+
\tilde{c}_{34}^{E}(\omega)\dtwo{\eps_{4t}}{z}+\nonumber\\
&&\qquad{}+\tilde{c}_{44}^{E}(\omega)\dtwo{\eps_{3t}}{y}+
\tilde{c}_{34}^{E}(\omega)\frac{\partial^2\eps_{3t}}{\partial
y\partial z}+
\tilde{c}_{33}^{E}(\omega)\dtwo{\eps_{3t}}{z},\nonumber\\
&&-\rho\omega^2\eps_{4t}=\tilde{c}_{24}^{E}(\omega)\dtwo{\eps_{2t}}{y}
+[\tilde{c}_{22}^{E}(\omega)+\tilde{c}_{23}^{E}(\omega)]\frac{\partial^2\eps_{2t}}{\partial
y\partial z}+
[2\tilde{c}_{24}^{E}(\omega)+\tilde{c}_{34}^{E}(\omega)]\dtwo{\eps_{2t}}{z}+\nonumber\\
&&{}\qquad+[\tilde{c}_{24}^{E}(\omega)+2\tilde{c}_{34}^{E}(\omega)]\dtwo{\eps_{3t}}{y}
+[\tilde{c}_{23}^{E}(\omega)+\tilde{c}_{33}^{E}(\omega)]\frac{\partial^2\eps_{3t}}{\partial
y\partial z}+ \tilde{c}_{34}^{E}(\omega)\dtwo{\eps_{3t}}{z}+
\nonumber\\
&& \label{fullsystem} \qquad{}+
\tilde{c}_{44}^{E}(\omega)[\dtwo{\eps_{4t}}{y}+\dtwo{\eps_{4t}}{z}].
\end{eqnarray}

The boundary conditions for $\eps_{it}$ follow from the assumption
that the crystal is simply supported, that is, it is traction free
at its edges (at $y=0$, $y=L_y$, $z=0$, $z=L_z$, to be denoted as
$\Sigma$)
\begin{equation}
\label{boundary1}
\sigma_1|_\Sigma=\sigma_2|_\Sigma=\sigma_3|_\Sigma=\sigma_4|_\Sigma=0.
\end{equation}
In our previous consideration \cite{ourrs2} this condition was
fulfilled at the corners of the crystal plate only, not at all its
edges. Eventually that led to an incorrect expression for the
resonance frequencies.

Substituting (\ref{boundary1}) into the constitutive relations
(\ref{2.4af}) and using (\ref{xi-sigma-eps}), we obtain the
boundary conditions for the strains in the following form
\begin{equation}\label{boundary2}
\eps_{it}|_\Sigma\equiv\eps_{i0}=d_{1i}(\omega)E_{1t},
\end{equation}
where
\begin{equation}
\label{d_om} d_{1i}(\omega) =  \sum\limits_{j = 1}^4
s_{ij}^E(\omega) e_{1j}(\omega), \end{equation} $e_{1i}(\omega)$
is given by (\ref{e_om}), and $s_{ij}^E(\omega)$ are the elements
of a matrix inverse to $\tilde{c}_{ij}^E(\omega)$ determined in
(\ref{c_om}).

Let us consider first the case of the paraelectric phases (at
$\tilde \xi=0$). Then
\begin{equation}
\label{boundary-para} \eps_{20}=\eps_{30}=0,
\end{equation}
as expected from the symmetry considerations. As we shall show
later, from this boundary condition it follows that
\[
\eps_{2t}(y,z)=\eps_{3t}(y,z)=0,
\]
at all $(y,z)$. Then the system (\ref{fullsystem}) reduces to a
single equation for $\eps_{4t}(y,z)$
\begin{eqnarray}
&& \label{singlesystem} -\rho\omega^2\eps_{4t}=
\tilde{c}_{44}^{E}(\omega)[\dtwo{\eps_{4t}}{y}+\dtwo{\eps_{4t}}{z}].
\end{eqnarray}
Its solution is
\begin{eqnarray}
&&\eps_{4t}(y,z)=\eps_{40}+\\
&&+\eps_{40}\sum_{k,l=0}^\infty
\frac{16}{(2k+1)(2l+1)\pi^2}\frac{\omega^2}{(\omega_{kl}^{(4)})^2-\omega^2}\sin\frac{\pi(2k+1)y}{L_y}\sin\frac{\pi(2l+1)z}{L_z},
\nonumber
\end{eqnarray}
with $\omega^{(4)}_{kl}$ given by
\begin{equation}
\label{res4} \omega_{kl}^{(4)}=\sqrt{\frac{\tilde
c_{44}^E(\omega^{(4)}_{kl})\pi^2}{\rho}\left[\frac{(2k+1)^2}{L_y^2}+\frac{(2l+1)^2}{L_z^2}\right]}.
\end{equation}
Since in the paraelectric phases
\[
\eps_{40}=\frac{e_{14}(\omega)}{\tilde{c}_{44}^E(\omega)}E_{1t}=d_{14}(\omega)E_{1t},
\]
we have that
\[
\overline{\eps_{4t}}=R_4(\omega)d_{14}(\omega)E_{1t},
\]
where
\begin{equation}
{R_4(\omega)}=1+ \sum_{k,l=0}^\infty
\frac{64}{(2k+1)^2(2l+1)^2\pi^4}\frac{\omega^2}{(\omega_{kl}^{(4)})^2-\omega^2}.
\end{equation}
Thus the dielectric permittivity reads
 \begin{equation}
 \label{dyn_susc}
\eps_{11}^\sigma(\omega)=\eps_{11}^{\eps}(\omega) +4\pi
R_4(\omega)e_{14}(\omega)d_{14}(\omega).
\end{equation}

Let us analyze the above results. In the static limit
($\omega\to0$, $R_4(\omega)\to1$)  from (\ref{dyn_susc}) we obtain
the static permittivity of a free crystal (see \cite{our-comp});
in the high frequency limit ($\sum_{k,l=0}^\infty
{64}/[(2k+1)^2(2l+1)^2\pi^4]=1$, and $R_4(\omega)\to 0$) we get a
dynamic permittivity (\ref{chi_clamped}) of a mechanically clamped
crystal, exhibiting relaxational dispersion in the microwave
region. Thus, eq. (\ref{dyn_susc}) explicitly describes the effect
of crystal clamping by high-frequency electric field.

 In the intermediate frequency region, it has a resonance dispersion with
numerous peaks ar frequencies where ${\rm
Re}[R_4(\omega)]\to\infty$. In this frequency range we can neglect
the frequency dependence of  $\tilde c_{44}^E(\omega)$ and reduce
the equation for the resonance frequencies (\ref{res4}) to an
explicit expression by putting in it $\tilde c_{44}^E(\omega)\to
c_{44}^E$.

Comparing (\ref{res4}) to the expression obtained previously
\cite{ourrs2} for a square X-cut
\[
\omega_{k}=\frac{\pi(2k+1)}{L}\sqrt{\frac{\tilde
c_{44}^E(\omega_k)}{\rho}},
\]
we can see that the incorrectly set boundary conditions
\cite{ourrs2} led to the $\sqrt2$ times smaller lowest resonance
frequency than the correct one. However, the low and high
frequency limits of the permittivity \cite{ourrs2} (the static
value and the clamped values with the relaxational dispersion in
the microwave region) were correct.

Now we shall proceed to the case of the ferroelectric phase. The
system of second-order partial differential equations
(\ref{fullsystem}) will be solved numerically using the finite
element method. However, its main features, such as most of the
resonant frequencies, including the lowest one, can be obtained
analytically, if we take into account the fact that $c_{22}$,
$c_{33}$, $c_{44}^E$, $c_{23} \gg c_{24}^E$, $c_{34}^E$.
Neglecting in (\ref{fullsystem}) the terms proportional to
$c_{24}^E$, $c_{34}^E$ (in the paraelectric phases
$c_{24}^E=c_{34}^E=0$ exactly) we get
\begin{eqnarray}
&&-\rho\omega^2\eps_{2t}=\tilde{c}_{22}^{E}(\omega)\dtwo{\eps_{2t}}{y}
+\tilde{c}_{44}^{E}(\omega)\dtwo{\eps_{2t}}{z}+
[\tilde{c}_{44}^{E}(\omega)+\tilde{c}_{23}^{E}(\omega)]\frac{\partial^2\eps_{3t}}{\partial
y^2}
 ,\nonumber\\
&&-\rho\omega^2\eps_{3t}=[\tilde{c}_{23}^{E}(\omega)+\tilde{c}_{44}^{E}(\omega)]\dtwo{\eps_{2t}}{z}+
\tilde{c}_{44}^{E}(\omega)\dtwo{\eps_{3t}}{y}+
\tilde{c}_{33}^{E}(\omega)\dtwo{\eps_{3t}}{z},\nonumber\\
&&-\rho\omega^2\eps_{4t}=[\tilde{c}_{22}^{E}(\omega)+\tilde{c}_{23}^{E}(\omega)]\frac{\partial^2\eps_{2t}}{\partial
y\partial
z}+[\tilde{c}_{23}^{E}(\omega)+\tilde{c}_{33}^{E}(\omega)]\frac{\partial^2\eps_{3t}}{\partial
y\partial z}+
\nonumber\\
&& \label{shortsystem} \qquad{}+
\tilde{c}_{44}^{E}(\omega)[\dtwo{\eps_{4t}}{y}+\dtwo{\eps_{4t}}{z}].
\end{eqnarray}
The system is partially split, with the two first equations
depending on $\eps_{2t}$ and $\eps_{3t}$ only. We look for the
solutions in the form of series
\begin{eqnarray}
&&\eps_{it}=\eps_{i0}+E_{1t}\sum_{n_y n_z}D_i^{n_y
n_z}\sin\frac{\pi n_y y}{L_y}\sin\frac{\pi n_z z}{L_z},\quad
i=2,3,4 \label{series}.
\end{eqnarray}
It is easy to verify that the boundary conditions
(\ref{boundary2}) are satisfied.

Substituting (\ref{series}) into the two first equations of
(\ref{shortsystem}) and noting that we can write that
\[
\eps_{i0}=\eps_{i0}\sum_{kl}\frac{16}{\pi^2 (2k+1)
(2l+1)}\sin\frac{\pi (2k+1)y}{L_y}\sin\frac{\pi (2l+1) z}{L_z},
\]
we obtain
\begin{eqnarray}
&& D_2^{kl}=\frac{16}{\pi^2 (2k+1)
(2l+1)}\frac{1}{\Delta_{kl}^{(23)}(\omega)}\left[-d_{13}(\omega)(\tilde{c}_{23}^E(\omega)+\tilde{c}_{44}^E(\omega))\frac{\pi^2(2k+1)^2}{L_y^2}+\right.\nonumber\\
&&\qquad\qquad\left.
+d_{12}(\omega)\left(\tilde{c}_{44}^E(\omega)\frac{\pi^2(2k+1)^2}{L_y^2}+\tilde{c}_{33}^E(\omega)\frac{\pi^2(2l+1)^2}{L_z^2}-\rho
\omega^2\right) \right],\nonumber\\
&& D_3^{kl}=\frac{16}{\pi^2 (2k+1)
(2l+1)}\frac{1}{\Delta_{kl}^{(23)}(\omega)}\left[-d_{12}(\omega)(\tilde{c}_{23}^E(\omega)+\tilde{c}_{44}^E(\omega))\frac{\pi^2(2l+1)^2}{L_z^2}+\right.\nonumber\\
&&\qquad\qquad+\left.
d_{13}(\omega)\left(\tilde{c}_{22}^E(\omega)\frac{\pi^2(2k+1)^2}{L_y^2}+\tilde{c}_{44}^E(\omega)\frac{\pi^2(2l+1)^2}{L_z^2}-\rho
\omega^2\right) \right] .\label{d23}
\end{eqnarray}
with
\begin{eqnarray}
&&\Delta_{kl}^{(23)}(\omega)=[\tilde{c}_{22}^E(\omega)\frac{\pi^2(2k+1)^2}{L_y^2}+\tilde{c}_{44}^E(\omega)\frac{\pi^2(2l+1)^2}{L_z^2}-\rho\omega^2]\times\nonumber\\
&&\qquad\times
 [\tilde{c}_{44}^E(\omega)\frac{\pi^2(2k+1)^2}{L_y^2}+\tilde{c}_{33}^E(\omega)\frac{\pi^2(2l+1)^2}{L_z^2}-\rho\omega^2]-\nonumber\\
&&
\quad-(\tilde{c}_{23}^E(\omega)+\tilde{c}_{44}^E(\omega))^2\frac{\pi^2(2k+1)^2}{L_y^2}\frac{\pi^2(2l+1)^2}{L_z^2}.
\end{eqnarray}
Note that owing to the boundary conditions (\ref{boundary-para}),
in the paraelectric phases $D_2^{n_y n_z}=D_3^{n_y n_z}=0$. It
means that the extensional modes associated with the strains
$\eps_{2}$ and $\eps_3$ are not excited by the longitudinal field
$E_1$, which is expected from the symmetry considerations for the
orthorhombic point group of Rochelle salt. The same results are
obtained by the numerical finite element method calculations for
the complete system (\ref{fullsystem}).

Substituting the found $\eps_{2t}(y,z)$ and $\eps_{3t}(y,z)$ from
(\ref{series}) with (\ref{d23}) into the third equation of
(\ref{shortsystem}) and then reexpanding two first terms in its
right-hand-side in series over $\sin\frac{\pi n_y
y}{L_y}\sin\frac{\pi n_z z}{L_z}$, we obtain that
\begin{equation}
\label{d4}
D_4^{kl}=\left[\frac{16d_{14}(\omega)}{\pi^2(2k+1)(2l+1)}+\sum_{mn}\frac{8}{\pi^2}Q_{mn}p_{km}p_{ln}\right]
\frac{\omega^2}{(\omega^{(4)}_{kl})^2-\omega^2},
\end{equation}
with
\begin{eqnarray}
&&
Q_{mn}=[\tilde{c}_{22}^E(\omega)+\tilde{c}_{23}^E(\omega)]D_2^{mn}+[\tilde{c}_{33}^E(\omega)+\tilde{c}_{23}^E(\omega)]D_3^{mn},
\end{eqnarray}
and \[ p_{km}=\frac{1}{m+k+1}
\]
if $k$ and $m$ are of the same parity, and
\[ p_{km}=\frac{1}{m-k}
\]
otherwise.

The averaged over the sample volume strains occurring in the
expression for the permittivity are then equal
\begin{equation}
\overline{\eps_{it}}=R_i(\omega)d_{1i}(\omega)E_{1t}
\end{equation}
with
\begin{equation}
R_i(\omega)=1+\sum_{k,l}\frac{4D^{kl}_i}{\pi^2(2k+1)(2l+1)d_{1i}(\omega)}
\end{equation}
and $D_i^{kl}$ given by (\ref{d23}) and (\ref{d4}). Finally, the
permittivity is
\begin{equation}
\eps_{11}^\sigma(\omega)=\eps_{11}^\eps(\omega)+4\pi\sum_{i=2,3,4}e_{1i}(\omega)d_{1i}(\omega)R_i(\omega).
\end{equation}

We are in a position now to determine the resonant frequencies of
the permittivity in the ferroelectric phase. Those occur at $
R_i(\omega)\to\infty $ and are of two types. The first type of
resonances is given by $R_{2,3}(\omega)\to\infty$, that is by
equation
\begin{equation}
\label{res23} \Delta_{kl}^{(23)}(\omega)=0,
\end{equation}
and it is associated with the extensional modes of $\eps_2$ and
$\eps_3$.  Solutions of (\ref{res23}), to be denoted as
$\omega_{kl}^{(23)}$, exist at all temperatures, but the
corresponding modes are not excited in the paraelectric phases.
Therefore, these resonances are present in the ferroelectric phase
only. On the other hand, $R_4(\omega)\to\infty$ both at
$\omega=\omega^{(23)}_{kl}$ and at $\omega=\omega^{(4)}_{kl}$
given by (\ref{res4}). The resonances given by (\ref{res4})
originate from the shear vibrational mode associated with $\eps_4$
and persist in the paraelectric phases. Such a division is,
however, artificial, as the modes are coupled. All three strains
calculated numerically from the complete system (\ref{fullsystem})
have resonances at the same frequencies.

\section{Numerical analysis}

The set of the model parameters, providing a fair description of
dielectric, piezoelectric, and elastic characteristics of Rochelle
salt, its microwave permittivity, as well as thermal expansion of
the crystal and the effects of hydrostatic and uniaxial pressures
has been obtained in \cite{our-diagonal,our-comp}. No additional
theory parameters need to be determined apart from those. However,
the sample dimensions should be specified. In the present paper we
shall use $L_y=1.60$~cm, $L_z=2.45$~cm of the Rochelle salt X-cut
sample, for which experimental data on the resonant frequencies
and the dielectric permittivity in the resonance region are
available \cite{Leonovici}. The static (equilibrium) values of the
dynamic variables $\tilde \xi$, $\tilde \sigma$, $\tilde \eps_i$
are calculated by minimization of the thermodynamic potential
(\ref{pot}) with respect to $\tilde\xi$, maximization with respect
to $\tilde \sigma$, and from equations (\ref{2.4a}).
Eqs.~(\ref{fullsystem}) for the strains were solved numerically
with the finite element method package \verb"FreeFem++"
\cite{freefem}. The solutions were used to find
$\overline{\eps_{it}}$ and, hence, the permittivity.

\begin{figure}[htb]
\centerline{\includegraphics[width=\textwidth]{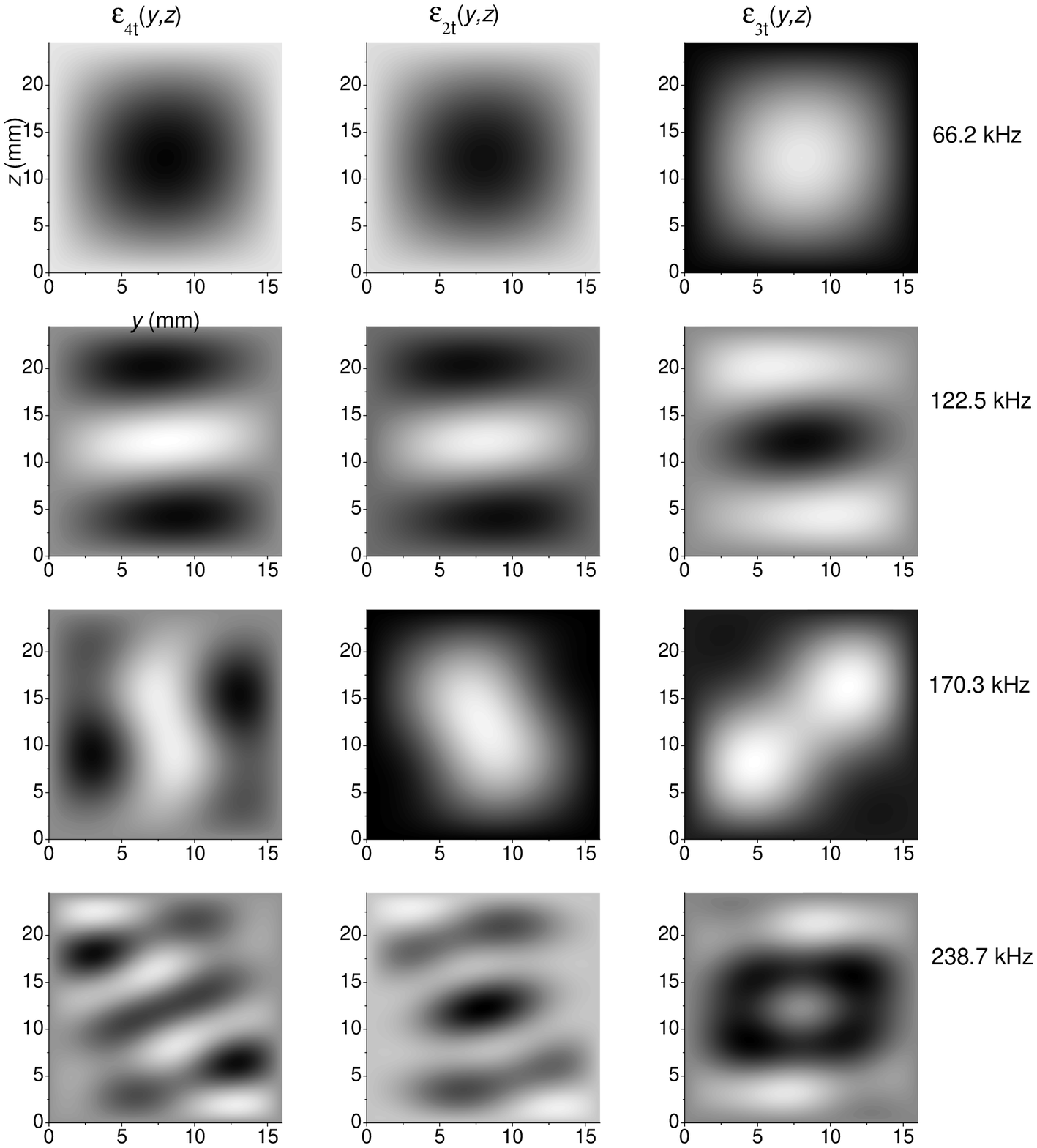}}
\caption{The distributions of the fluctuational parts of the
strains $\eps_{it}(y,z)/E_{1t}$ at different frequencies of a
X-cut of Rochelle salt with $L_y=1.60$~cm, $L_z=2.45$~cm at 293~K.
} \label{eps-yz1}
\end{figure}

\begin{figure}[htb]
\centerline{\includegraphics[width=\textwidth]{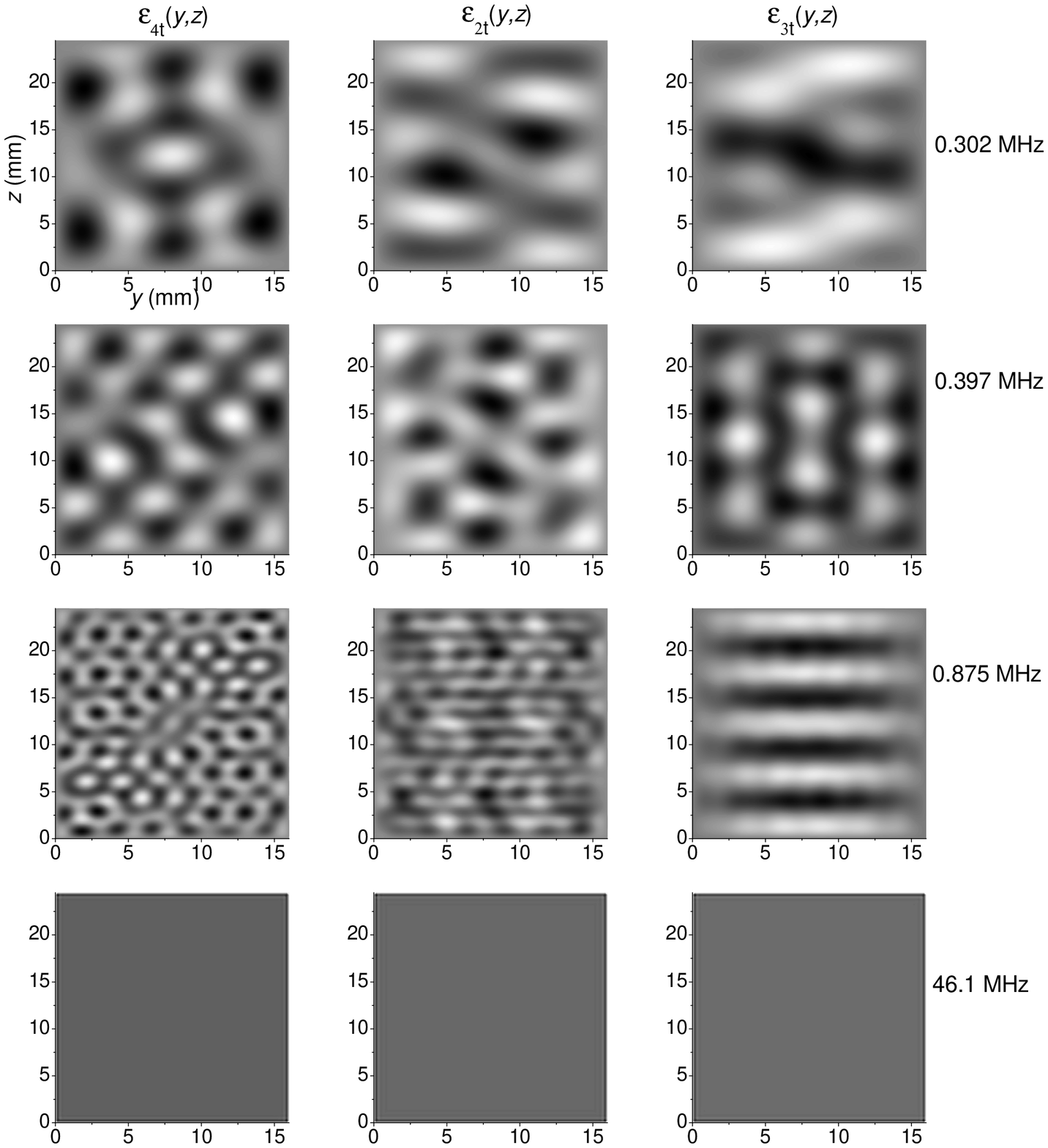}}
\caption{The same. } \label{eps-yz2}
\end{figure}

In figures~\ref{eps-yz1} and \ref{eps-yz2} we show the spatial
distribution of the fluctuational parts of the strains
$\eps_{it}(y,z)/E_{1t}$ at different frequencies, obtained by
solving the full system (\ref{fullsystem}) with the boundary
conditions (\ref{boundary2}). Note that the gray scales are
different for each graph.


The distributions have a single extremum at the sample center at
low frequencies (up to the frequency of the first resonance); then
the extrema multiply. Above the resonances, $\eps_{it}(y,z)$ are
zeros in most of the plate, only going to the boundary values
given by (\ref{boundary2}) within very narrow strips near the
sample edges. It illustrates the effect of crystal clamping by a
high-frequency electric field.

\begin{figure}[htb]
\centerline{\includegraphics[width=0.5\textwidth]{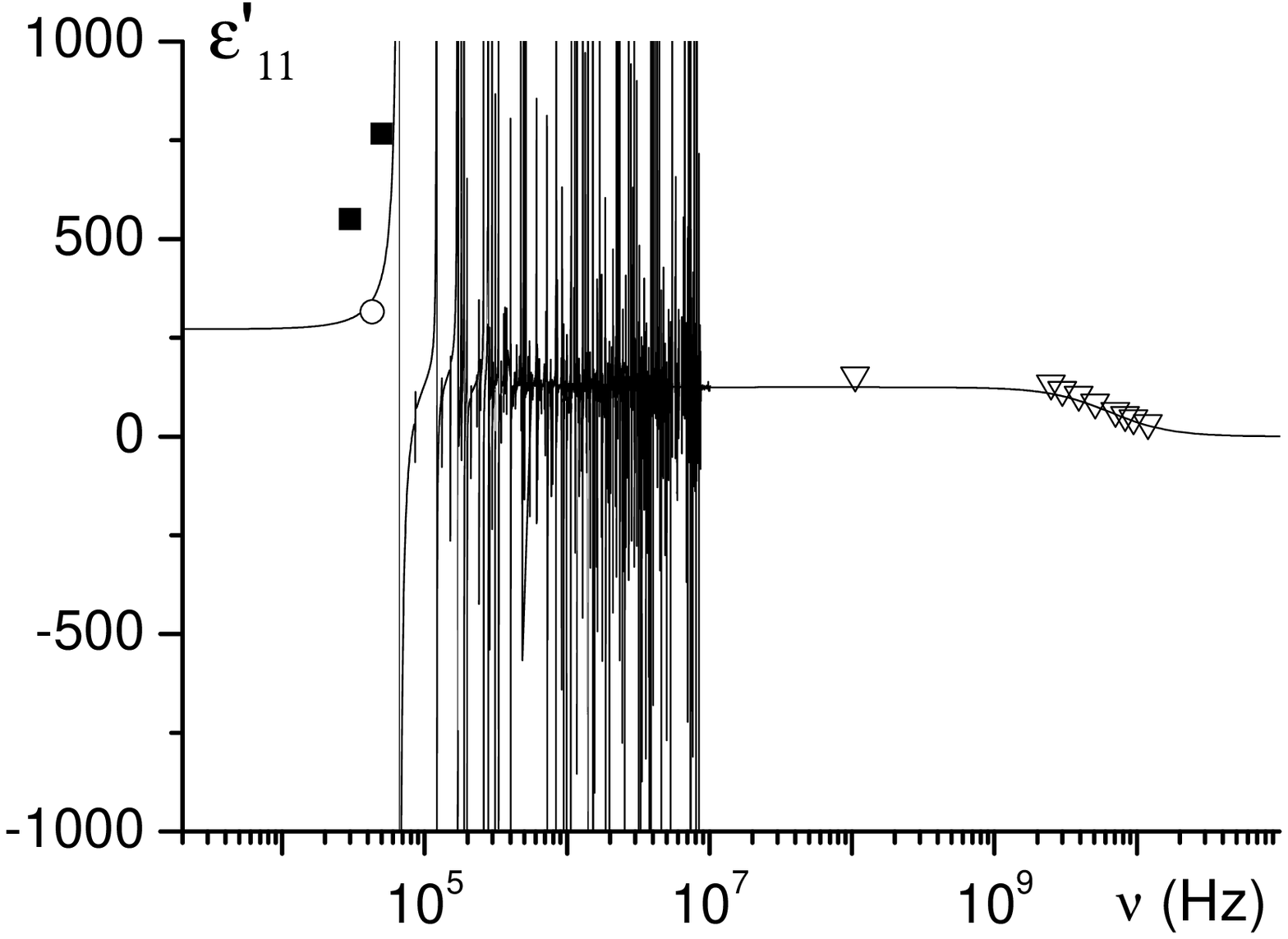}}
\caption{Frequency dependence of dielectric permittivity of a
X-cut of Rochelle salt at 293~K. Symbols are experimental points
taken from \cite{Leonovici} -- $\blacksquare$, \cite{int4} --
$\bigtriangledown$, \cite{Lunk} -- $\circ$. The line: a theory.
The line and $\blacksquare$ are for $L_y=1.60$~cm, $L_z=2.45$~cm.
Other symbols correspond to samples of different sizes. }
\label{eps-n}
\end{figure}

Figure~\ref{eps-n} shows the frequency dependence of dynamic
permittivity of the Rochelle salt X-cut (with $L_y=1.60$~cm,
$L_z=2.45$~cm) within the entire frequency range of the current
model applicability.  That range does not include the region of
the domain-related dispersion below 1~kHz \cite{Poplavko1,Araujo}
or the submillimeter (100-700~GHz) region of resonant dispersion
\cite{Volkov}. The experimental data shown by open symbols are for
the frequencies outside the piezoelectric resonance regions of the
samples used in the measurements, so the dimensions of those
samples are irrelevant. The obtained evolution of the permittivity
is analogous to the experimental \cite{Poplavko1} and to the
previously obtained theoretical \cite{ourrs2,Andrusyk} ones: from
the static permittivity of a free crystal at low frequencies, via
the piezoelectric resonance region ($10^4\div10^7$~Hz for this
sample dimensions), to the clamped crystal value, and, eventually,
to a relaxational dispersion in the microwave region. A fairly
good agreement with experiment is obtained.

\begin{figure}[htb]
\centerline{\includegraphics[width=\textwidth]{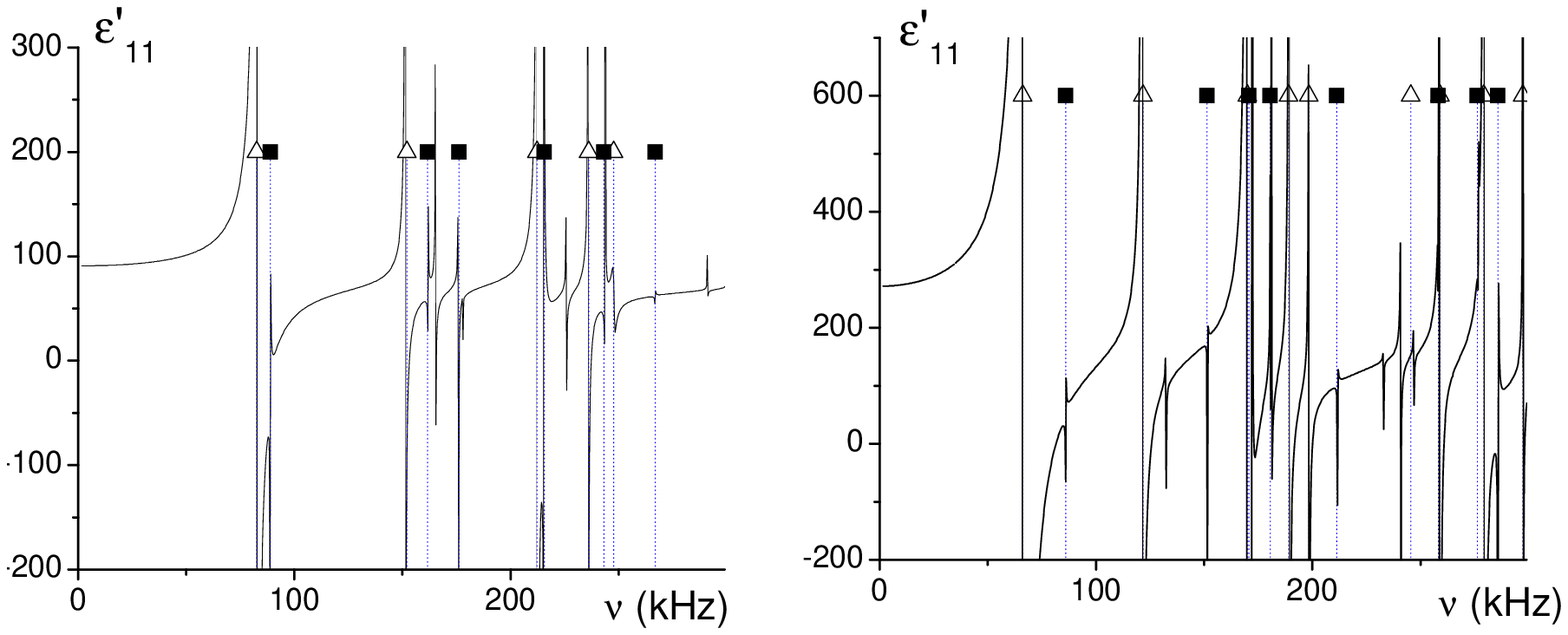}}
\caption{Frequency dependence of the dynamic dielectric
permittivity of a Rochelle salt X-cut  at 275 K (left) and 293 K
(right). The solid line is calculated with $\overline{\eps_{it}}$
found from (\ref{fullsystem}). $\triangle$ and $\blacksquare$ are
the resonant frequencies $\nu_{kl}^{(4)}$  and $\nu_{kl}^{(23)}$
of the simplified system (\ref{shortsystem}). $L_y=1.60$~cm,
$L_z=2.45$~cm.} \label{eps-f}
\end{figure}

Now we take a closer look at the resonance region.  Ability of the
simplified system (\ref{shortsystem}) to describe the resonant
behavior of lattice strains and, henceforth, the dynamic
dielectric permittivity of Rochelle salt X-cuts is demonstrated in
fig.~\ref{eps-f}, showing the frequency dependence of the
permittivity in the lower part of the resonant region. One can see
that most of the resonant frequencies of the permittivity,
calculated numerically, using the complete set of equations
(\ref{fullsystem}), are very well reproduced by the resonant
frequencies of the simplified system (\ref{shortsystem}).

\begin{figure}[htb]
\centerline{\includegraphics[width=0.45\textwidth]{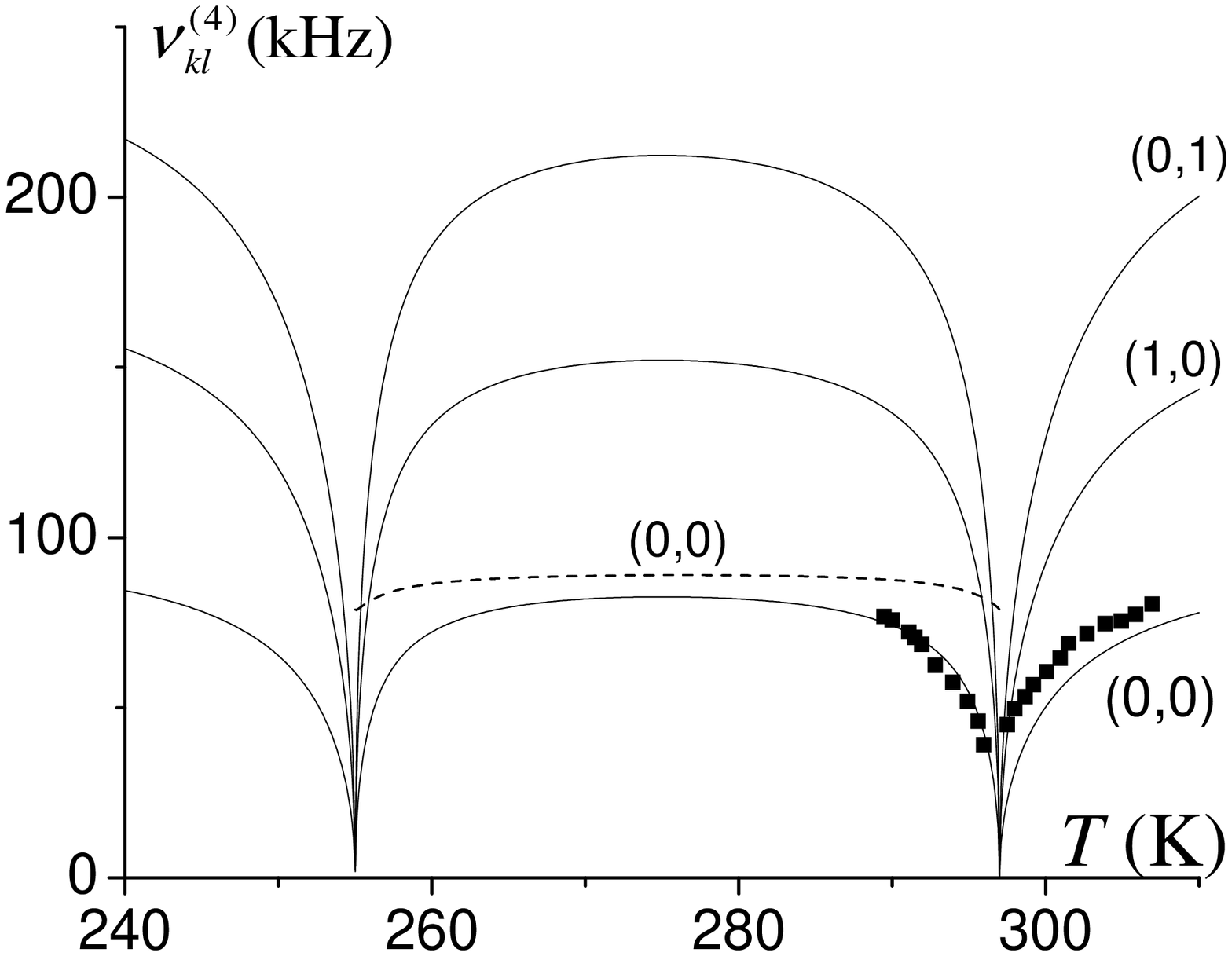}~~~~~~~
\includegraphics[width=0.45\textwidth]{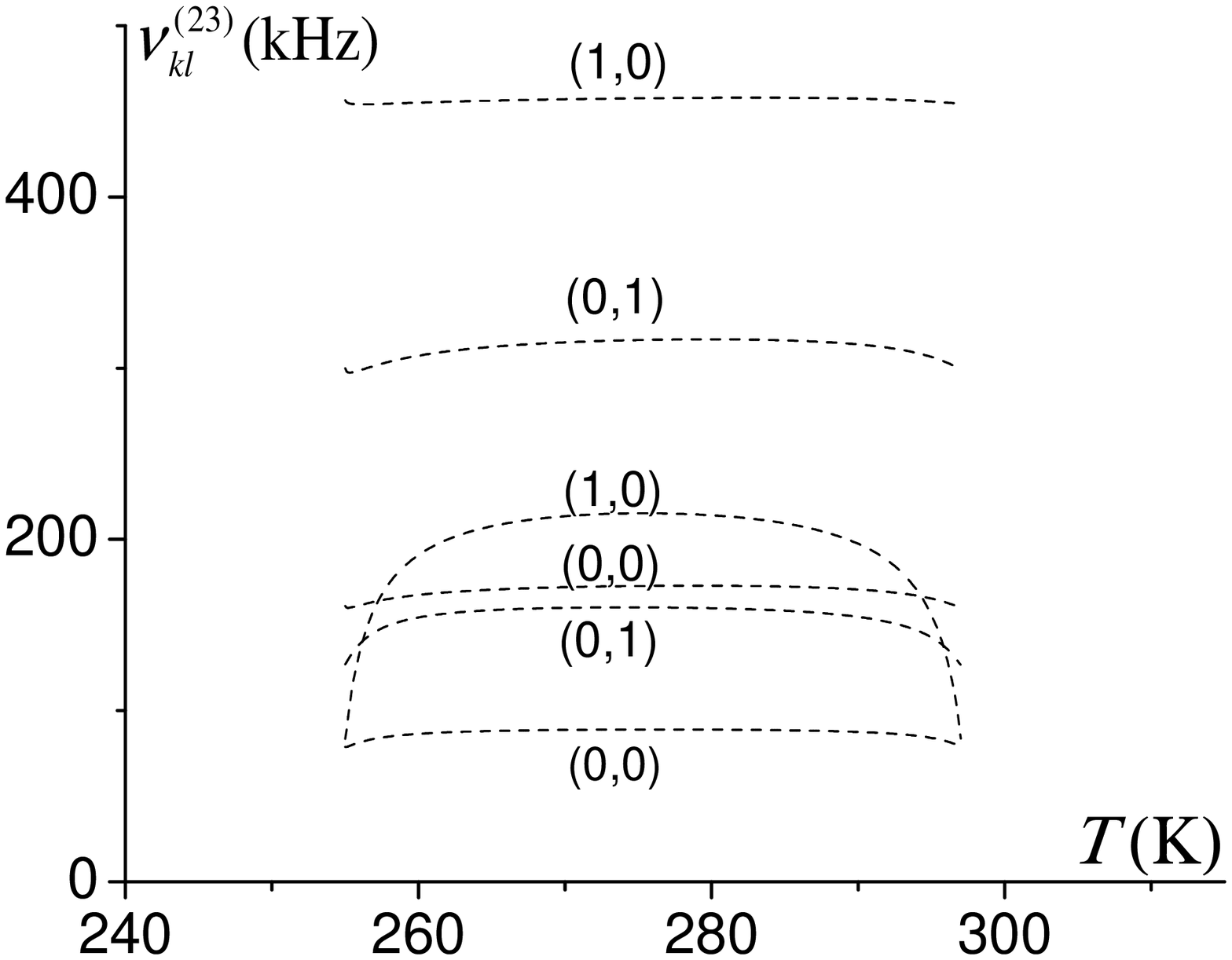}
}
\caption{Temperature dependence of the lowest resonance
frequencies  of the Rochelle salt X-cut with $L_y=1.60$~cm,
$L_z=2.45$~cm. Lines: the theory. Solid lines: $\nu_{kl}^{(4)}$;
dashed lines: $\nu_{kl}^{(23)}$. Symbols: experimental points of
\cite{Leonovici}. The numbers in parentheses are the $(k,l)$
values.} \label{lowres}
\end{figure}

The temperature dependence of a few lowest resonance frequencies
of the simplified system (\ref{shortsystem}) for this particular
X-cut is shown in figure~\ref{lowres}. It is seen that at all
temperatures the lowest is the resonance frequency
$\nu_{kl}^{(4)}=\omega_{kl}^{(4)}/2\pi$ associated with the shear
mode at $k=l=0$. The lowest frequency
$\nu_{kl}^{(23)}=\omega_{kl}^{(23)}/2\pi$ of the extensional modes
(at $k=l=0$), active in the ferroelectric phase only, is higher at
any temperature and at any sample dimensions and always remains
finite. The shear mode frequencies $\nu_{kl}^{(4)}$, on the other
hand, go to zero at
the Curie temperatures at all $k$ and $l$.

The agreement with experiment for the lowest resonant frequency is
quite good in the ferroelectric phase and gets worse in the upper
paraelectric phase. It is, apparently, caused by a similar misfit
for the elastic constant $c_{44}^E$ (see Eq.~(\ref{res4}) and
\cite{our-comp}).

\begin{figure}[hbt]
\centerline{\includegraphics[height=0.22\textheight,width=0.48\textwidth]{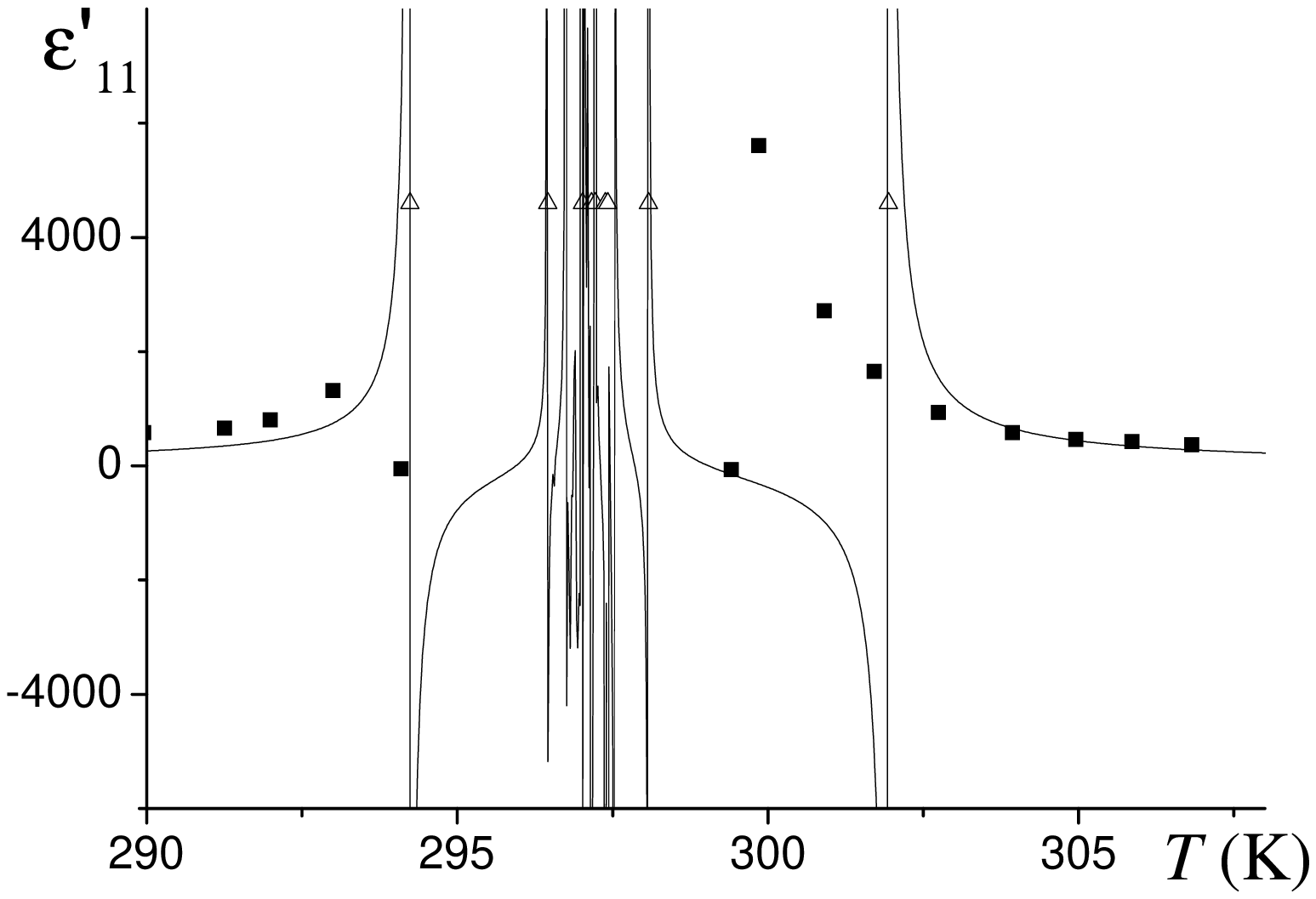}~~~~
\includegraphics[height=0.22\textheight,width=0.48\textwidth]{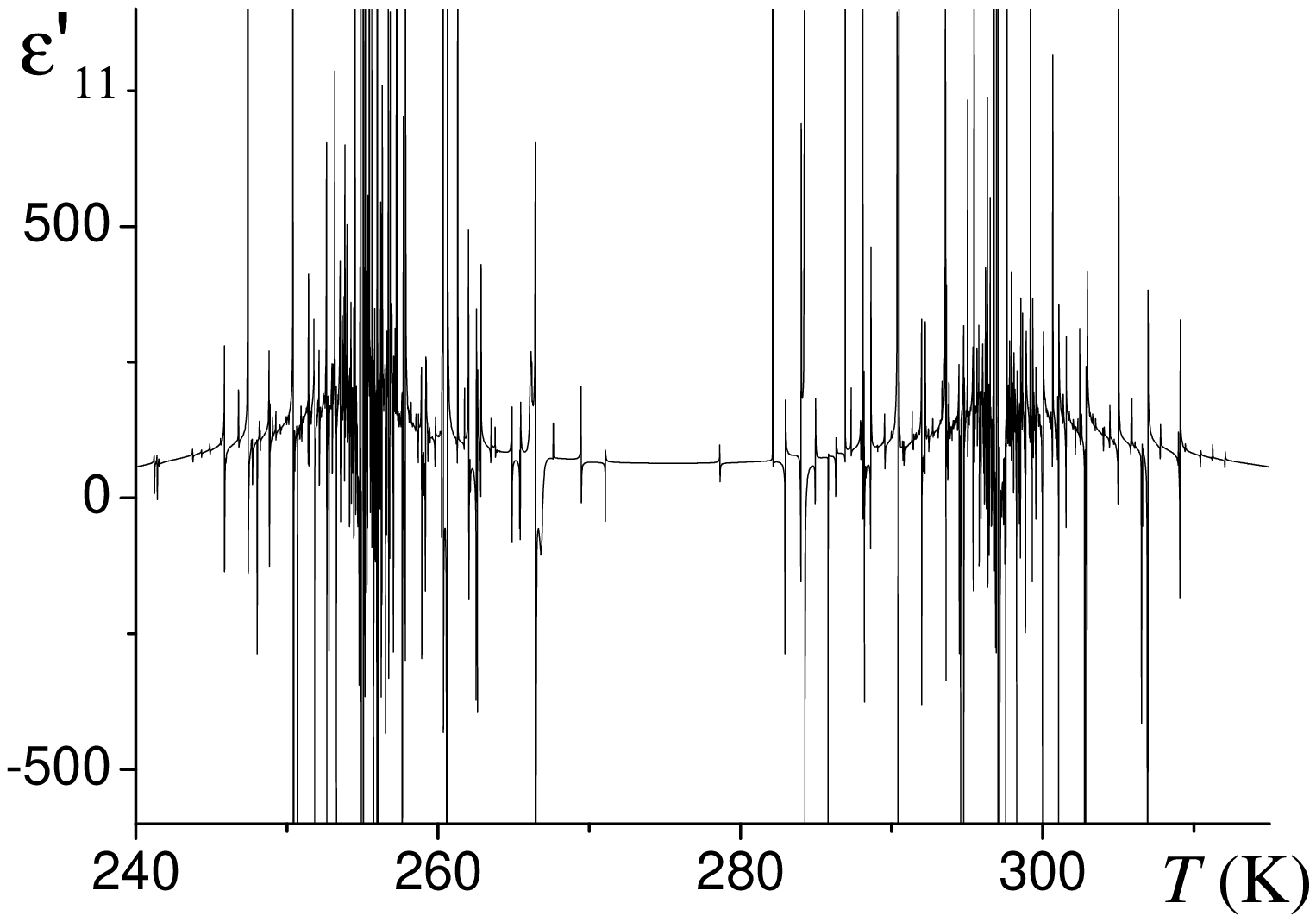}}
\caption{Temperature dependences of dynamic dielectric
permittivity of the Rochelle salt X-cut with $L_y=1.60$~cm,
$L_z=2.45$~cm  at 60~kHz (left) and 800~kHz (right).  Solid line:
the theory. $\bullet$: experimental points of \cite{Leonovici};
$\triangle$ are the resonant frequencies of the simplified system
(\ref{shortsystem}) given by (\ref{res4}).} \label{eps-t}
\end{figure}

This disagreement is well revealed in the temperature variation of
the dynamic dielectric permittivity of the Rochelle salt X-cut at
60~kHz shown in fig.~\ref{eps-t} (left). There are two distinct
and well resolved resonance peaks below and above the Curie point.
With increasing frequency, these peaks move away from the Curie
point. They are associated with the $\eps_4$ shear mode and given
by $\nu_{00}^{(4)}$. In the upper paraelectric phase, the
disagreement between theory and experiment for $\nu_{res}$ seen in
fig.~\ref{lowres} resulted in a more than 2~K difference between
the theoretical and experimental temperatures of the resonant
peak.

What has not been observed experimentally \cite{Leonovici} is that
apart from the two discussed peaks, in the close vicinity of the
transition temperature, the permittivity also has a multitude of
other resonances, which are higher order resonances of the
$\eps_4$ mode ($\nu_{kl}^{(4)}$ at $k+l>0$).

Actually, it follows from fig.~\ref{lowres} that at frequencies
below the lowest resonant frequency of the extensional mode $
\nu^{(23)}_{00}(T_{\rm C})$ at the Curie points  (the threshold
frequency which can be found from (\ref{res23}) at $c_{44}^E\to
0$; it equals 80~kHz for this particular X-cut), all resonances
are associated with the shear $\eps_4$ mode. Above
$\nu^{(23)}_{00}(T_{\rm C})$, the resonances associated  with the
extensional  modes appear in the ferroelectric phase.

The temperature curve of the permittivity in a wider temperature
range shows (fig.~\ref{eps-t}, right) that the resonances form two
packs around each Curie temperature, with the density of
resonances increasing at approaching the Curie points. It can be
shown that the pack widths increase with increasing frequency,
which is caused by the shape of the $\nu_{kl+}^{(4)}(T)$ curves
(see fig.~\ref{lowres}) and by appearance of the $\nu_{kl}^{(23)}$
resonances. Eventually, at some sufficiently high frequency the
packs overlap.

\section{Concluding remarks}

Vibrations of X-cuts of Rochelle salt crystals and their influence
on the dynamic dielectric permittivity are analyzed using the
modified Mitsui model that takes into account the shear strain
$\eps_4$ and the diagonal strains $\eps_1$, $\eps_2$, $\eps_3$
\cite{our-diagonal}. The system dynamics is described within the
frequency range, starting from 1~kHz (above the dispersion
associated with the domain wall motion) via the piezoelectric
resonance region and the microwave relaxational dispersion up to
about $10^{11}$~Hz.

Special attention is paid to the piezoelectric resonance region.
Explicit expressions for the resonant frequencies, associated with
the shear mode of $\eps_4$ and with the extensional  in-plane
modes of $\eps_2$, $\eps_3$, of such cuts are derived. They are
obtained at neglecting the out-of-plane mode associated with
$\eps_1$, as well as the elastic constants $c^E_{24}$ and
$c^E_{34}$. The temperature behavior of the resonant frequencies
is analyzed; it is shown that the lowest resonance is associated
with the shear mode at all temperatures.

The changes in the calculated spatial distributions of the strains
with increasing frequency visualize the effect of crystal clamping
by the high-frequency electric field. Both the shear mode and the
extensional modes are suppressed.

It is shown that the resonances associated with the extensional
modes appear above a certain threshold frequency, and in the
ferroelectric phase only, which is consistent with the symmetry
considerations.  In the close vicinities of the transition
temperatures, the permittivity has a multitude of overlapping
peaks, which are higher order resonances of the $\eps_4$ mode.






\begin{thebibliography}{100}


\bibitem{Araujo}
J.F. Araujo, J. Mendes Filho et al, Phys. Rev. B {\bf 57}, 783
(1998).

\bibitem{shylnikov}
A.V. Shyl'nikov, N.M. Galijarova et al. Kristallografiya {\bf 31},
326 (1986).

\bibitem{Poplavko1}
Y.M.Poplavko, V.V.Meriakri, P.Pereverzeva, V.V.Alesheckin, and
V.I.Molchanov, Fiz. Tverd. Tela (Leningrad) {\bf 15}, 2515 (1974)
 [Sov. Phys. Solid State {\bf 15}, 1672 (1974).


\bibitem{Leonovici}
M.R.~Leonovici and I.~Bunget, Ferroelectrics {\bf 22}, 835 (1979).

\bibitem{Mason}
W.P.Mason,  Phys. Rev.  {\bf 55}, 775 (1939).

\bibitem{muller}
H.~Mueller, Phys. Rev. {\bf 58}, 565 (1940).


\bibitem{int4}
F.~Sandy and  R.V.~Jones,  Phys. Rev. {\bf 168}, 481 (1968).


\bibitem{Volkov}
A.A. Volkov, G.V. Kozlov, S.P. Lebedev, JETP {\bf 52}, 722 (1980).


 \bibitem{int3}
 T. Mitsui,  Phys. Rev. {\bf  111}, 1259 (1958).

\bibitem{83}
B. Zeks,  G.C. Shukla, R.Blinc, Phys. Rev. B. {\bf  3}, 2306
(1971).

\bibitem{86}
 B.Zeks,  G.C.Shukla,  R.Blinc, J. Phys. {\bf 33},
supplement to N 4, c2-67-c2-68 (1972).

\bibitem{our-comp-piezo}
R.R.~Levitskii, I.R.Zachek, T.M.Verkholyak, A.P.Moina,
 Condens. Matter Phys. \textbf{6},  261 (2003).

\bibitem{ourrs2}
A.P.~Moina, R.R.Levitskii and I.R.Zachek, Phys. Rev.~B
\textbf{71}, 134108 (2005).

\bibitem{ourrs}
R.R.~Levitskii, I.R.~Zachek, T.M.~Verkholyak, and A.P.~Moina,
Phys. Rev.~B {\bf 67}, 174112 (2003).

\bibitem{Andrusyk}
A. Andrusyk (2011). Piezoelectric Effect in Rochelle Salt,
Ferroelectrics - Physical Effects, Mickael Lallart (Ed.), InTech,
http://www.intechopen.com/articles/show/title/piezoelectric-effect-in-rochelle-salt

\bibitem{our-diagonal}
A.P.~Moina, R.R.Levitskii and I.R.Zachek, Condens. Matter Phys.
\textbf{14}, 43602 (2011).

\bibitem{int9}
R.J. Glauber, J. Math. Phys. {\bf 4}, 294 (1963).

\bibitem{Masonbook}
W.P.~Mason, {\it Piezoelectric Crystals and Their Application to
Ultrasonics} (Van Nostrand, New York, 1950).



%
%
%












\bibitem{our-comp}
A.P.Moina,  Condens. Matter Phys. \textbf{15} (2012). (to be
published)





\bibitem{freefem} http://www.freefem.org/ff++






\bibitem{Lunk}
U. Schneider, P. Lunkenheimer, J. Hemberger,  and A. Loidl,
Ferroelectrics, \textbf{242}, 71 (2000).








 \end{thebibliography}



\end{document}